# Thermal properties of SmFeAs(O$_{1-x}$F$_x$) as probe of the interplay between electrons and phonons

M.Tropeano,[1] A.Martinelli,[2] A.Palenzona,[2] E.Bellingeri,[1] E.Galleani d'Agliano,[1] T.D.Nguyen,[3] M. Affronte[3] and M.Putti[1]

[1]*CNR-INFM-LAMIA and Dipartimento di Fisica, Via Dodecaneso 33, 16146 Genova, Italy*
[2]*CNR-INFM-LAMIA and Dipartimento di Chimica and chimica Industriale, Via Dodecaneso 31, 16146 Genova, Italy*
[3]*CNR-INFM-S3 and Dipartimento di Fisica, Università di Modena e Reggio Emilia Via G.Campi 213/A, I-41100 Modena, Italy*

A comparative study of thermal properties of SmFeAsO, SmFeAs(O$_{0.93}$F$_{0.07}$) and SmFeAs(O$_{0.85}$F$_{0.15}$) samples is presented. Specific heat and thermal conductivity show clear evidences of the spin density wave (SDW) ordering below T$_{SDW}$~135 K in undoped SmFeAsO. At low level of F-doping, SmFeAs(O$_{0.93}$F$_{0.07}$), SDW ordering is suppressed and superconducting features are not yet optimally developed in both specific heat and thermal conductivity. At optimal level of F-doping SmFeAs(O$_{0.85}$F$_{0.15}$) anomalies related to the superconducting transition are well noticeable. By a compared analysis of doped and undoped samples we conclude that, despite F-doping modifies definitely the electronic ground state, it does not substantially alter phonon and electron parameters, like phonon modes, Sommerfeld coefficient, electro-phonon coupling. The analysis of the thermal conductivity curves provides an evaluation of SDW and superconducting energy gap, showing that phonons can suitably probe features of electronic ground state.

## 1. Introduction

The recent discovery of superconductivity at critical temperature T$_c$ up to 50 K in layered rare-earth (*RE*) iron based oxipnictide compounds *RE*FeAsO (*RE*=La, Ce, Pr, Nd, Sm)[1,2,3,4] has sparked enormous interest in this class of materials. Like high-T$_c$ cuprates they have layered structure with conducting FeAs layers sandwiched between insulating *RE*O layers and exhibit superconductivity at relatively high temperatures upon electron[1,2,3,4,5] or hole doping[6] of the related non superconducting compounds. The latter exhibit a magnetic transition with a sharp drop of resistivity below 140-150K.[1,2,3] This anomaly has been explained within controversial theoretically frameworks.[7,8,9,10,11,12,13,14] Optical measurements[7] suggest that LaFeAsO has an antiferromagnetic spin-density-wave (SDW) state. This has been confirmed by neutron scattering studies[15] which provide evidences for an antiferromagnetic long-range ordering with a small 0.35μB per Fe moment below 140K. Nearly at the same temperature a structural transition have been reported for LaFeAsO,[15] NdFeAsO [16] and recently for SmFeAsO.[17]

More recently a new class of double layer AFe$_2$As$_2$ (A=Sr, Ba, Eu) compounds have been investigated. They present even more pronounced resistivity and specific heat anomalies at a temperature ranging from 130 to 200 K and T$_c$ increases up to 38 K when doped with holes.[18,19,20,21]

First principle calculations indicate that electron-phonon coupling is not sufficient to explain superconductivity.[8,22] On the other hand, thermal properties are strongly affected by the establishment of SWD[7,18] suggesting that these properties can be useful tools to investigate details of electronic structure. To our knowledge, an in-dept analysis of thermal properties of oxipnictides is still lacking in the literature.

In this work we report results of specific heat and thermal conductivity measurements on SmFeAsO, which exhibits anomalies at ~135 K. Although the nature of this anomaly is not yet definitively clear, in the following we refer to it as due to transition to a spin density wave (SDW) ordered state. The same properties measured on SmFeAs(O$_{0.93}$F$_{0.07}$) show that anomalies at ~135K are suppressed while the superconducting features are still not well developed. Superconducting anomalies are clearly evident in both specific heat and thermal conductivity of SmFeAs(O$_{0.85}$F$_{0.15}$). A lambda anomaly is clearly evident in the specific heat of the three sample close to liquid helium temperature. Remarkably, the lattice properties are very close in the three samples and this allows direct comparison of the thermal properties of doped and undoped cases leading to an evaluation of SDW and superconducting energy gap. It turns out from this analysis that thermal properties can suitably probe details of the electronic ground state. For sake of clarity, we focus on the thermal properties of the parent SmFeAsO in comparison with the SmFeAs(O$_{0.93}$F$_{0.07}$) sample first and we discuss the superconducting properties of SmFeAs(O$_{0.85}$F$_{0.15}$) secondly.

## 2. Experimental Details

The SmFeAsO, SmFeAs(O$_{0.93}$F$_{0.07}$) and SmFeAs(O$_{0.85}$F$_{0.15}$) samples were prepared in three steps as described in ref. 23: 1) synthesis of SmAs from pure elements; 2) synthesis of SmFeAsO and SmFeAs(O$_{1-x}$F$_x$) by reacting SmAs with stoichiometric amounts of Fe, Fe$_2$O$_3$ and FeF$_2$ at high temperature; 3) grinding of the so obtained sample and further sintering at high temperature, in order to obtain a compact sample. The samples were characterised by X-ray powder diffraction followed by Rietveld refinement, revealing their single-phase nature. TEM analysis evidences the lack of structural defects. The effect of sintering is to increase density and connection between the grains improving substantially transport properties.

Heat capacity and thermal conductivity were measured by Quantum Design PPMS. Heat capacity measurements were performed by using the two-tau method.

Resistivity and magnetization measurements reported elsewhere,[23,24] evidence a pronounced anomaly around 140 K in SmFeAsO specimen. SmFeAs(O$_{0.93}$F$_{0.07}$) does not reveal SDW anomaly and shows a rather broad superconducting transition below T$_c$~34 K probably related to somewhat inhomogeneous distribution of fluorine. SmFeAs(O$_{0.85}$F$_{0.15}$), instead, undergoes a sharp superconducting transition at T$_c$=51.5 K as detected by resistivity and susceptibility measurements[24] and exhibits well developed and sharp anomalies in the thermal properties as discussed below.

## 3. The SDW State

*3.1 Specific Heat.*

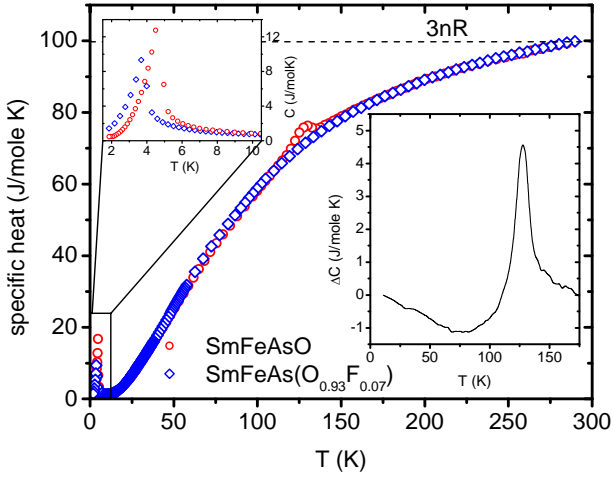

**Figure 1** Temperature dependence of the specific heat of SmOFeAs and SmO$_{0.93}$F$_{0.07}$FeAs. Upper inset: specific heat anomaly $\Delta C(T)$ related to the SDW.

The temperature dependence of the specific heat $C$ of SmFeAsO is plotted in fig.1 and it can be compared with that of SmFeAs(O$_{0.93}$F$_{0.07}$) and with that SmFeAs(O$_{0.85}$F$_{0.15}$) in fig.5. Remarkably, the absolute values of $C$ are quite close for the three samples. At high temperature, the C values tends to saturate to the Dulong Petit value 3nR with n=4 and R=8.314J/molK the gas constant. These features lead to the conclusion that fluorine doping induces only little quantitative changes in the lattice as expected from the almost identical crystalline structure.[17] This result is fully consistent by phonon density of states measurements in undoped and optimally doped LaFeAsO$_{1-x}$F$_x$.[25] Our results thus rectify data previously reported by Ding *et al.* [26] for which the specific heat of SmFeAsO$_{1-x}$F$_x$ largely exceeded the Dulong Petit value and strongly depended on F doping.

The main difference between the two $C(T)$ curves in fig.1 is the anomaly clearly visible and peaked at about 130 K in the undoped SmFeAsO sample but not in SmFeAsO$_{0.93}$F$_{0.07}$. This anomaly was previously observed[7,26] and ascribed to the SDW transition. We can better show this anomaly by evaluating $\Delta C(T) = C(T)_{SmFeAsO} - C(T)_{SmFeAs(OF)}$. This is plotted in the inset of fig.1. Its cusp shape suggests an important role of fluctuations above T$_{SDW}$~130K and its height is about 5 J/molK. $\Delta C(T)$ becomes negative below 110 K. If we assume that $C_{SmFeAs(OF)}$ approximates the normal state (no SDW order) behaviour, qualitatively this implies the entropy difference between the normal and the gapped states tends to be compensated at $T_{SDW}$ like in the superconducting transition.

Both SmFeAsO and SmFeAs(O$_{0.93}$F$_{0.07}$) compounds show sharp peaks in $C(T)$ at 4.6 K and 3.7 K respetively (see magnification in fig. 1). In SmFeAs(O$_{0.85}$F$_{0.15}$) this peak is a bit smeared. Above these peaks, the $C(T)$ curves fit well a dependence $C/T = \gamma + \beta T^2$. In spite of some uncertainty due to the presence of the peaks, the following parameters can be obtained by fitting data between 15K and 25K:

$\gamma=(42\pm2)$ mJ/molK$^2$ and $\beta=(0.36\pm0.04)$ mJ/molK$^4$ for SmFeAsO, $\gamma=(44\pm2)$ mJ/molK$^2$ and $\beta=(0.35\pm0.04)$ mJ/molK$^4$ for SmFeAs(O$_{0.93}$F$_{0.07}$) and $\gamma=(43\pm2)$ mJ/molK$^2$ and $\beta=(0.36\pm0.04)$ mJ/molK$^4$ for SmFeAs(O$_{0.85}$F$_{0.15}$). It is worth noting that the γ coefficients are similar in doped and undoped samples and are higher (about one order of magnitude) than the Sommerfeld coefficient for the parent compound LaFeAsO.[7,27] Similar value of the γ coefficients were reported for (Sm$_{1.85}$Ce$_{0.15}$)CuO$_{4-\delta}$.[28] which presents similar layered structure and the Sm$^{3+}$ sublattice ordinates exactly at the same temperature (4.7K) The entropy removal related to these peaks can be evaluated as $S_m = \int C_m/T \cdot dT$ where $C_m=C-(\gamma T+\beta T^3)$. For both SmFeAsO and SmFeAs(O$_{0.93}$F$_{0.07}$), $S_m$ tends to saturate to Rln2 (see the inset of figure 2) as expected for a doublet ground state of Sm$^{3+}$. This leads us to conclude that the low temperature peaks can be actually related to the AFM transition of the whole Sm$^{3+}$ sublattice with some –little-sensitivity to the electron doping. Within this framework, the relatively high γ values might indicate that some hybridization/interaction of the electron wave functions with the Sm$^{3+}$ magnetic ion may lead to renormalization of the effective electron mass. This is supported by resistivity measurements which show a drop in correspondence of Sm$^{3+}$ AFM transition[23] and by the rather high value of Pauli susceptibility.[24] However, magnetic excitations related to the incipient antiferromagnetic transition of the Sm$^{3+}$ sublattice may also contribute to such high γ value.

The β coefficients are reasonably close for the two compounds and a bit lower than those evaluated for LaFeAsO.[12] In the low temperature limit, acoustic phonon branches are expected to characterize the lattice vibrations and the β coefficient can be related to the Debye temperature through the relation $\beta = \frac{12}{5}\pi^4 R(T/\Theta_D)^3$ from which we can estimate $\Theta_D$=175 K for SmFeAsO and SmFeAs(O$_{0.85}$F$_{0.15}$), $\Theta_D$=173 K for SmFeAs(O$_{0.93}$F$_{0.07}$), respectively. This rather low $\Theta_D$ values well agree with acoustic modes as evaluated in ref. 22.

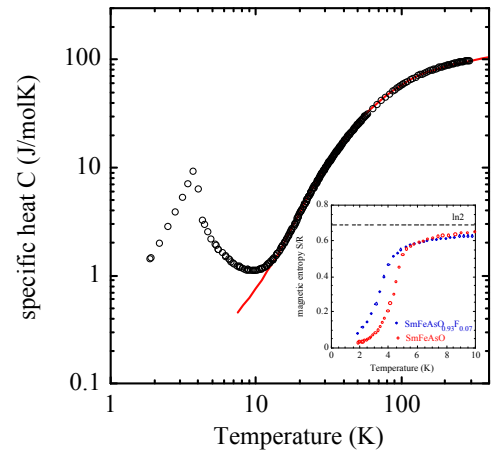

**Figure2** Specific heat data of SmFeAsO$_{0.93}$F$_{0.07}$ and $C(T) = \gamma T + C_D(T) + C_E(T)$ (continuous line). Inset: S$_m$ associated to the ordering of the Sm$^{3+}$ sublattice.

At higher temperature (T≥30K) optical modes contribute as well to the specific heat. By considering optical modes centered at 100 cm$^{-1}$, 180 cm$^{-1}$ and 290 cm$^{-1}$ as evaluated in ref. 22, the overall high temperature $C(T)$ behavior can be obtained by considering Einstein contributions, $C_E$, in addition to the Debye, $C_D$ and the Sommerfeld contributions. This is shown in figure 2 where specific heat of SmFeAsO$_{0.93}$F$_{0.07}$ above 10 K is well reproduced by the sum of the three contributions $C(T) = \gamma T + C_D(T) + C_E(T)$.

*3.2 Thermal Conductivity*

In figure 3, the thermal conductivity κ of SmFeAsO and SmFeAs(O$_{0.93}$F$_{0.07}$) is plotted as a function of temperature. Undoped sample presents a clear signature of SDW transition, abruptly increasing below $T_{SDW}$ ~135 K. Similar behaviour was observed in LaFeAsO [29], yet the κ values in our case are more than two times higher indicating a good crystallinity of the sample. Measurements at B=9T yield a κ(T) curve, not shown for sake of clarity, that perfectly overlaps data in zero field. This indicates that in undoped sample mechanisms involved in heat conduction are essentially insensitive to magnetic field.

SmFeAs(O$_{0.93}$F$_{0.07}$) sample shows thermal conductivity values smaller than those of SmFeAsO and very different behaviour. In agreement with specific heat measurements there is no feature around 135 K. Instead maximum occurs around 20 K, as previously observed in ref. [30]. The application of magnetic field reduces thermal conductivity removing this anomaly. A close data inspection (see the inset) shows that κ(9T) departs from κ(0T) below ~34 K, which roughly corresponds to the superconducting critical temperature of this sample.[23]

In LaFeAsO compounds[29,30] it was observed that thermal conductivity is dominated by phonons. This is true also for our samples: by evaluating the electronic contribution κ$_e$(T) by the Wiedemann Franz law ($\kappa_e = L_0 T/\rho$ where $L_0$=2.45×10$^{-8}$ WΩ/K and $\rho$ is the resistivity) it actually turns out that κ$_e$(T)/κ(T) is less than 1% for SmFeAsO and less than 10% for SmFeAs(O$_{0.93}$F$_{0.07}$).

From these evaluations it emerges that, independently of the rare earth and of sample quality, the heat conduction is dominated by phonons in particular in sample with low carrier density. Within this framework, an abrupt rise of κ is expected in correspondence to the gap opening at the Fermi surface followed by carrier condensation and consequent suppression of electron-phonon scattering.

This explains the features observed in figure 3: In SmFeAsO, κ rises below $T_{SDW}$ and in SmFeAs(O$_{0.93}$F$_{0.07}$) and in SmFeAs(O$_{0.85}$F$_{0.15}$) (this will shown in the next section) below T$_c$. In the case of SmFeAs(O$_{0.93}$F$_{0.07}$) the application of magnetic field, producing pair breaking, removes nearly completely the anomaly (see the inset of fig.3).

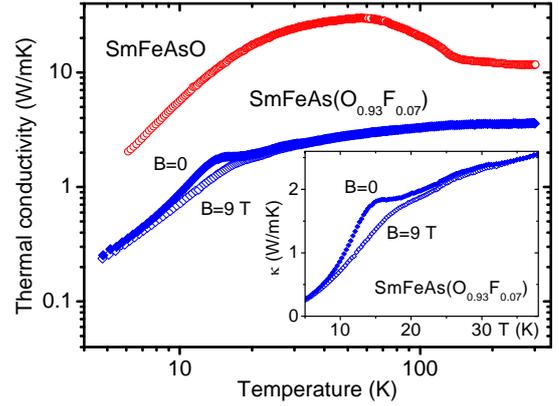

**Figure 3** κ of SmFeAsO and SmFeAs O$_{0.93}$F$_{0.07}$ as a function of temperature. The inset shows the magnification of the superconducting transition at B=0 and B= 9 T.

For a more quantitative analysis, a phenomenological model that describes the phonon thermal conductivity of superconductors can be used. The model developed initially by Bardeen, Rickayzen and Tewordt (BRT) in the framework of the BCS theory [31] was later generalized[32] and gives the following expression for the phonon thermal conductivity:[33]

$$\kappa_{ph} = A t_D^3 \int_0^{1/t_D} dx \frac{e^x x^4}{(e^x-1)^2} \tau(t_D, x, y) \quad (1)$$

where $t_D = T/\Theta_D$, $x = h\nu/k_B T$ and $y = \Delta(T)/k_B T$ are the reduced temperature, phonon energy and gap energy. The parameter A, assuming that only longitudinal acoustic phonons contribute to κ$_{ph}$, is approximately:32

$$A = \left(\frac{4}{3}\pi\right)^{1/3} \frac{k_B^2 L_b}{h a^2} \Theta_D \quad (2)$$

where $a$ is an average lattice constant and $L_b$ is the typical grain size. $\tau(t_D, x, y)$ is the normalised relaxation time that, according to the Mathiessen rule, may be written as:32

$$\tau(t_D, x.y)^{-1} = \left[1 + \alpha t_D^4 x^4 + \chi t_D x g(x,y) + \eta t_D^4 x^2\right] \quad (3)$$

The terms proportional to α, χ and η refer to the main phonon relaxation rates: Point-defects, charge carriers and other phonons, respectively. Each relaxation rate is divided by the phonon relaxation time with sample boundaries $\tau_b^{-1} = v_s/L_b$ ($v_s$ is the sound velocity in the materials). The function $g(x, y)$, which is included in the electron-phonon term, is the ratio between the electron-phonon scattering times in the normal and in the superconducting state and has been derived in the original BRT theory.[32] The strong similarities existing between superconducting and SDW ground states[34] suggest that the same derivation can be applied here, introducing as free parameter $\sigma_{SDW}=\Delta_{SDW}(0)/k_B T_{SDW}$ and assuming a BCS temperature dependence of the gap.

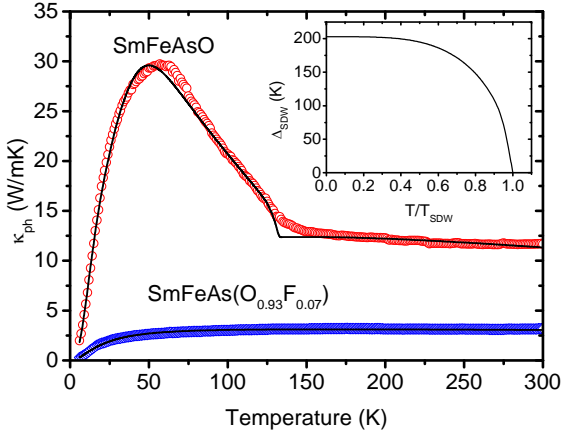

**Figure 4** $\kappa_{ph}$ of SmFeAsO and SmFeAs O$_{0.93}$F$_{0.07}$ as a function of temperature. Continuous line are the best fitting curves evaluated by eq. (1) with the parameter in table I.

In figure 4 the phonon contributions to the thermal conductivity, $\kappa_{ph}$, of SmFeAsO and SmFeAs(O$_{0.93}$F$_{0.07}$), are plotted as a function of temperature. $\kappa_{ph}$ is evaluated as: $\kappa_{ph} = \kappa - \kappa_e \approx \kappa - L_0 T / \rho$.

In the case of SmFeAs(O$_{0.93}$F$_{0.07}$), neglecting the superconducting transition, $\kappa_{ph}$ has been evaluated in the normal state by keeping $\kappa$(B=9T) and by extrapolating $\kappa_e$ linearly to zero in the superconducting state. Eq. (1) depends on the following set of free parameters: A, $\Theta_D$, $\alpha$, $\chi$, $\eta$ and $\sigma_{SDW}$. In order to introduce some constraints, the two curves are simultaneously fitted by keeping A, $\Theta_D$ and $\eta$ equal for the two samples. $\Theta_D$ and $\eta$ are indeed related to phonon spectrum which is not affected by doping as shown by specific heat measurements and A is related to the grain size which is similar in the two samples.23 The best fitting curves are plotted in figure 4 as continuous lines. They capture the main features giving a quite satisfying agreement. In the case of SmFeAsO the increase below $T_{SDW}$ is well reproduced even if the fluctuations which make less stiff the curve around the transition are not taken into account. The best fitting parameters are reported in Table I.

**Table 1**: Best fit parameters of the thermal conductivity of SmFeAsO and SmFeAs(O$_{0.93}$F$_{0.07}$)

| sample | A (W/mK) | $\Theta_D$ (K) | $\eta$ | $\alpha$ | $\chi$ | $\sigma_{SDW}$ |
|---|---|---|---|---|---|---|
| SmFeAsO | 0.19(1) | 178(2) | 9(1) | 149(2) | 27 (2) | 2.90(2) |
| SmFeAs(O$_{0.93}$F$_{0.07}$) | | | | 2240(40) | 29 (2) | - |

$\Theta_D$ essentially coincides with the value obtained by low temperature specific heat. By the A parameter an average grain length L$_b$ ~6 μm can be evaluated by eq. (2), assuming an average lattice constant a=5 Å: this well agrees with grain size reported in ref.23. The different absolute values of the two curves are mainly taken into account by the parameter $\alpha$ which is more then one order of magnitude higher in SmFeAs(O$_{0.93}$F$_{0.07}$) than in SmFeAsO. This coefficient is proportional to point defect scattering which is strongly enhanced by F substitutions. The parameter $\chi$ is nearly the same in the two samples. For free electrons $\chi$ is proportional to the density of state at the Fermi level,[32] $N(0)$, thus suggesting that $N(0)$ is substantially unchanged after doping. The same can be concluded by comparing the above reported Sommerfeld $\gamma$ constant of the two samples. The coefficient $\chi$ can be expressed in terms of the longitudinal acoustic phonon contribution to the electron-phonon coupling constant, $\lambda_{la}$:[32] $\chi \approx (\pi/2)(k_B T_{SDW}/\bar{t})(L_b/a)\lambda_{la}$ where $\bar{t} \approx N(0)^{-1}$ is the effective hopping matrix element for a two-dimensional tight-binding band of electrons. With $T_{SDW}$=135 K, L$_b$=6 μm, a=5Å, and N(0)~2.6-2.1 states/eV[9,22] we obtain $\lambda_{la}$~0.04-0.05. This value fairly agrees with the estimated contribution of the longitudinal acoustic mode to the total electron-phonon coupling constant as inferred from ref.22.

Finally we consider the $\sigma_{SDW}$ coefficient which is expected to assume the BCS value 3.52. In our case we evaluated $\sigma_{SDW}=2\Delta_{SDW}(0)/k_B T_{SDW}$ =2.9. In the inset of figure 4 the temperature dependence of SDW gap with $\Delta_{SDW}(0)$= 199 K=17 meV is plotted.

Within a BCS framework the specific heat anomaly can be evaluated as $\Delta C_{SDW}/\gamma T_{SDW} \approx 1.43(\sigma_{SDW}/3.52)$, where the height of the jump has been scaled with the reduced gap. Assuming $\gamma$=42 mJ/mol K, $T_{SDW}$=135 K, $\sigma_{SDW}$=2.9, we obtain $\Delta C_{SDW}$~6.6 J/mol K, not far from the value of 5 J/mol K experimentally evaluated.

## 4. Superconducting State.

*4.1 Specific Heat*

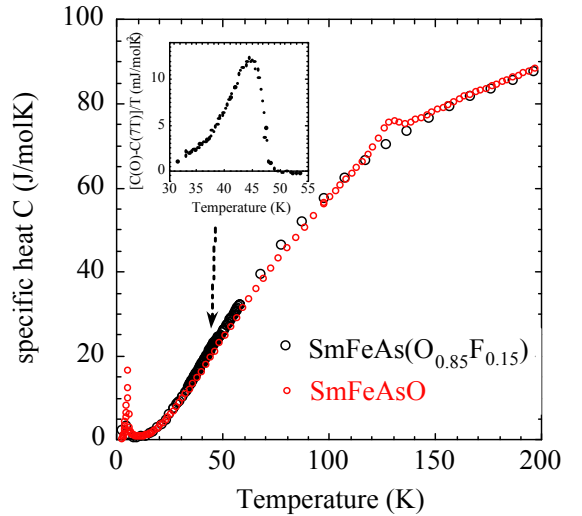

**Figure 5** Specific heat of SmFeAs(O$_{0.85}$F$_{0.15}$) (black circles) as compared with that of SmFeAsO (red circles).

Fig. 5 shows the specific heat of SmFeAsO$_{0.85}$F$_{0.15}$ as compared to that of undoped SmFeAsO. The two *C(T)* curves essentially overlap each other except in proximity of the respective anomalies: while SmFeAsO show the SDW bump at 130K, the specific heat of SmFeAs(O$_{0.85}$F$_{0.15}$) exhibits a small jump at about 47K. By subtracting the specific heat *C(7T)* measured in 7T applied magnetic field, we plot this specific heat jump as [*C(0T)-C(T)*]/T in the inset of fig.5. The jump is relatively sharp and starts at 47K and the value at its maximum is [*C(0T)-C(T)*]/T=12mJ/molK$^2$. This is almost twice the values previously reported.[26,35] but it is still less than 0.1% of the absolute specific heat value. This explains why in not optimally doped and/or not homogeneous samples like SmFeAs(O$_{0.93}$F$_{0.07}$) the anomaly was not observed.[27] Now if we use the $\gamma$ = 43 mJ/molK$^2$, it results [*C(0T)-C(T)*]/$\gamma$T=0.27, i.e. a value much smaller than what expected from BCS. This can be partially due to the small inhomogeneity within the sample but it may also indicate that the normal state electronic contribution is over-estimated and $\gamma$ also comprises contribution from spin excitations related to the antiferromagnetic transition of Sm$^{3+}$ sublattice as previously mentioned.

*4.2 Thermal Conductivity*

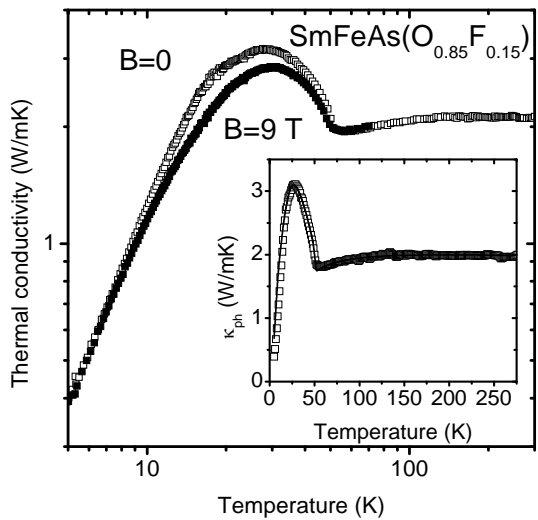

Fig. 6 shows the thermal conductivity κ of SmFeAs(O$_{0.85}$F$_{0.15}$) as a function of temperature at B=0 and B= 9 T. κ is rather constant at high temperature and abruptly increases below 51 K showing a maximum around 30 K. The application of the magnetic field reduces κ, but the maximum remains well evident. Following step by step the discussion done in the previous section we evaluate that the electron contribution on the thermal conductivity is less than the 10%. Thus, also at this level of doping the heat conduction is dominated by phonons and the rise of κ is due to electron condensation below T$_c$. Differently from the case of SmFeAs(O$_{0.93}$F$_{0.07}$) the application of magnetic field of 9 T, does not remove the anomaly, indicating that such a field is much lower than the upper critical field in this sample.

The thermal conductivity of this sample is analyzed within the same theoretical frame introduced in section 3.2. κ$_{ph}$ versus temperature is plotted in the inset of fig. 6 where continuous line is the best fitt evaluated by eq. (1) with the parameters listed in table II it well reproduces the experimental data with a reasonable set of parameters. A, $\Theta_D$ and ν essentially coincide with the values reported in table I for SmFeAsO and SmFeAs(O$_{0.93}$F$_{0.07}$) as expected for similar microstructure and specific heat values of the three samples. $\alpha$ is nearly two times higher than the value of SmFeAs(O$_{0.93}$F$_{0.07}$), as expected for a double fluorine content. χ is 30% lower than the values of SmFeAsO and SmFeAs(O$_{0.93}$F$_{0.07}$), suggesting a decreasing of *N(0)* with increasing doping. Finally the reduced gap value is 2Δ/k$_B$T$_c$=3.9±0.3 from which we evaluate Δ(0)=8.5±1 meV. This result is in good agreement with the value obtained by Andreev Spectroscopy on SmFeAs(O$_{0.85}$F$_{0.15}$).[36] We point out, however, that our evaluation assumes a single BCS gap and it does not consider the occurrence of different pairing symmetries or more energy gaps as recently suggested by some authors.[27,37,38,39,40,41] The inclusion of more complex gap structures in the model could in principle improve the quality of the fit, mainly at low temperature, but it is beyond the purpose of this work.

**Figure 6** Thermal conductivity of SmFeAs(O$_{0.85}$F$_{0.15}$) as a function of temperature at B=0 and B= 9 T. The inset shows κ$_{ph}$ vs T. Continuous line is the best fit evaluated by eq. (1) with the parameters in table II.

**Table II**: Best fit parameters of the thermal conductivity of SmFeAs(O$_{0.85}$F$_{0.15}$)

| sample | A (W/mK) | $\Theta_D$ (K) | η | α | χ | T$_c$ (K) | σ=2Δ(0)/k$_B$T$_c$ |
|---|---|---|---|---|---|---|---|
| SmFeAs(O$_{0.85}$F$_{0.15}$) | 0.13 (1) | 180(2) | 10(2) | 4300(300) | 20(2) | 51.5 | 3.9(3) |

### 5. Conclusions

From the comparative analysis of thermal properties of undoped and F-doped SmFeAsO two important hints on the nature of the SDW can be drawn:
1) The SDW transition, evident in both specific heat and thermal conductivity of SmFeAsO, disappears at low level of F doping with the occurrence of superconductivity. On the other hand, the phonon structure and, more noteworthy, the electronic density of states are not strongly affected by doping. These results follow by the observation that in the two samples the lattice specific heat is essentially unchanged, and also the Sommerfeld coefficients γ and the electron-phonon scattering rate parameters χ are nearly equal. This suggests that doping abruptly breaks the symmetries of the Fermi surface inhibiting the SDW formation in favor of superconductivity, without changing substantially both electron and phonon density of states.

2) The analysis of the thermal properties has emphasized similarities between SDW and a superconducting ground state. The difference between the specific heat of undoped and doped sample looks qualitatively like the difference between the superconducting and the normal specific heat even if a cusp shape suggests an important role of fluctuations. The thermal conductivity of undoped and optimally doped samples present similar abrupt rise below the respective ordering temperatures, $T_{SDW}$ and $T_c$, These behaviors can be well rationalized by describing the SDW transition within a BCS generalized model with $2\Delta_{SDW}(0)/k_BT_{SDW}$ =2.9 and the superconducting transition within a BCS single gap model with $2\Delta(0)/k_BT_c$ =3.9.

We conclude that electron-phonon coupling strongly characterizes the thermal properties giving important clues on the electronic ground states.

This work is partially supported by MIUR under the projects PRIN2006021741.